\documentclass{article}

\usepackage{preprint}
\usepackage{amsmath}
\usepackage{upgreek}				
\usepackage[utf8]{inputenc} 		
\usepackage[T1]{fontenc}    		
\usepackage[dvipsnames]{xcolor}	
\usepackage{hyperref}       		
\hypersetup{
    colorlinks = true,
    linkcolor = {NavyBlue},		
    anchorcolor = .,
    citecolor = {NavyBlue},		
    filecolor = .,				
    menucolor = .,
    runcolor = {cyan}, 
    urlcolor = {NavyBlue},		
    }
\usepackage{url}            
\usepackage{booktabs}       
\usepackage{amsfonts}       
\usepackage{nicefrac}       
\usepackage{microtype}      
\usepackage{fancyhdr}       
\usepackage{graphicx}       
\usepackage{setspace} 
\usepackage{soul} 
\usepackage{ragged2e} 
\usepackage[toc,page]{appendix}
\usepackage{longtable}
\usepackage{eso-pic}
\usepackage{tikz}
\usepackage[superscript]{cite}

\graphicspath{{./Figures/}}     
\newcommand{\beginsupplement}{%
        \setcounter{table}{0}
        \renewcommand{\thetable}{S\arabic{table}}%
        \setcounter{figure}{0}
        \renewcommand{\thefigure}{S\arabic{figure}}%
     }

  \centering
\title{On Electropolymerized Fingerprints and their Potential for Identification and Encryption}

\author{
	Antoine Baron\textsuperscript{a}, 
	Luc Brulin\textsuperscript{a}, 
	Corentin Scholaert\textsuperscript{a}, 		Yannick Coffinier\textsuperscript{a}, 		Fabien Alibart\textsuperscript{b},      \\
  \bf{S\'{e}bastien Pecqueur\textsuperscript{a}}\\
  \\
  a. IEMN, UMR 8520 \\
Univ. Lille, CNRS, Univ. Polytechnique Hauts-de-France\\
59000 Lille, France\\
  \\
  b. Laboratoire Nanotechnologies \& Nanosyst\`{e}mes (LN2)\\
  CNRS, Universit\'{e} de Sherbrooke\\
  J1X0A5, Sherbrooke, Canada\\
  \\
  \texttt{sebastien.pecqueur@iemn.fr} \\
}

\begin{document}
\maketitle

\begin{abstract}
While human technology is ruled by determinism, biological systems exploit a subtle balance of control and stochasticity. This balance, evident in the morphogenesis of textural patterns imprinted on leaves, fur or skin can help hierarchize organisms both as a representative of their species and as unique individuals. 
In this study, we identified that, by exploiting electrochemistry, it is possible to generate such versatile but specific textures, to imprint patterns of a conducting polymer on a conducting substrate. 
It is shown that the 1D morphogenesis of conducting polymer dendrites on wires translates, on 2D surfaces, as highly heterogeneous coatings of dark spots, rosettes or marbled patterns.
Despite their inherent stochasticity, these patterns are characteristic of the physical conditions they grew in, and particularly of the chemical content of the electroactive solution used for their electropolymerization.
A statistical study demonstrates that these patterns could be used as fingerprints to physically tag the identity of a solution within a specific class.
By the identification of a new electrochemical process which allows generating physical fingerprints with optical, electrical and chemical contrast on an electrode, this research paves the way toward a disruptive low-cost technology which could allow any end-user to generate personal tags on a glass slide or on a micro-chip, to engrave physically-encrypted personal information for various applications.

\end{abstract}

\raggedright
\keywords{conducting polymer \and cryptography \and electropolymerization \and electroconvection}

\newpage 

\justifying

\section{Introduction}

Whether when crafting a tool or engineering a complex system, human technology has mainly relied on deterministic design that rejects randomness.\cite{Li_2007,735728}
Natural systems, on the other hand, have developed optimized functional structures through stochastic mechanisms and direct interactions with their environment.\cite{Bejan_2006}
The emergence of such patterns and structures under physical and chemical forces is called morphogenesis.\cite{Turing1952}
This phenomenon may manifest in the two-dimensional distinct spot patterns on animal coats or fish skin,\cite{Nakamasu_2009,Liu2006} as well as in the three-dimensional ramified structures characteristic of diffusion-limited aggregation (DLA).\cite{Witten_1981}
DLA is mathematically governed by the Laplace equation, the steady-state limit of the diffusion. equation\cite{Paterson_1984} 
Diffusion is also involved in reaction-diffusion systems such as Turing models, which have been proposed to explain fingers and spot formation in biological systems.\cite{Nakamasu_2009,Raspopovic2014,Watanabe2012,Liu2006}
In fact, the fundamental transport phenomena driving natural morphogenesis ---reaction, diffusion, convection, and migration--- are also heavily involved in electrochemical processes.
This parallel suggests that by learning to control these growth phenomena, we could engineer energy-efficient systems that exploit natural stochasticity and self-organization to dynamically create low-energy structures throughout their lifetime using electrochemistry.\\
In laboratory settings, an effective method to achieve controlled and dynamic growth of such structures is by applying electrical voltages to drive chemical reactions that form solid conducting structures.\cite{Melrose_1992} 
This may be achieved through the electrodeposition of metals, or through electropolymerization specifically for polymers. 
Simulation and experimental results indicate that structures such as 2D Turing-like spot patterns can be controlled through precise parameters of the electrochemical deposition.\cite{Sgura2012,Mazouz2000, Li2001} 
Under far-from-equilibrium conditions, it is possible to grow 3D ramified structures from metals called dendrites,\cite{Chazalviel_1990,Bai_2016} which have presented parallels with plant growth,\cite{Fleury_1999} and with structures generated through DLA.\cite{Peklar_2024, Li_2024_dendrites}\\ 
Conducting polymers, can yield similar ramified structures under diffusion-controlled conditions.\cite{Kaufman_1987}
Several research groups have studied such polymers as the material for analog circuits,\cite{Koizumi2016, Scholaert2025, Janzakova2021,Baron_2025,Watanabe_2024} because of their peculiar ion/electron coupling.\cite{Tropp2023,Paulsen2019}
Their growth process involves an alternating high-voltage signal applied on anisotropic electrodes to trigger electropolymerization.\cite{Janzakova2021a}
By relaxing strict fabrication constraints, this method may allow matter to structure itself in response to an environment, reflecting the thermodynamic and chemical conditions of the growth medium.\\
A sometimes overlooked phenomenon in the study of pattern formation is the role of hydrodynamics through convection.
In nature, convective vortices have been observed driving the emergence of complex structures, whether by orchestrating the mechanical forces impacting tissue morphogenesis, as seen during the formation of chick embryos (3D),\cite{Fleury_2011,Cartwright_2008,Chuai_2009} or by governing thermal instabilities, such as with frost patterns on grass (2D).\cite{Ackerson_2015}
In electrochemistry, analogous vortices can form predictably upon the application of a sufficiently strong electric field between the electrodes during a phenomenon called electroconvection.
In fact, electroconvection has been previously mentioned as the possible cause for hexagonal pattern formation during a chemical reaction.\cite{Pashchanka_2021} 
Acting as a mechanical force, electroconvective vortices could dictate preferential deposition sites for reactants, thereby governing the final shape and the characteristic wavelength of the electrodeposit.
While electroconvection has also been identified as a critical factor in the growth of metallic dendrites,\cite{Huth_1995, Ma_2021, Rubinstein_2024} its role has not been explicitly reported in the context of electropolymerization reactions.\\
This work started with the observation that the electropolymerization of PEDOT:PSS, when conducted with a parallel-plates electrode geometry as shown in Fig.~\ref{fig:fig1}a, does not yield vertical dendrites (3D) but localized spots on the electrodes instead (2D, as pictured in Fig.~\ref{fig:fig1}b).
Considering a uniform electric field is expected in this geometry, a symmetry breaking instability must have occurred to lead to an alternation of dark and light spots depicted in Fig.~\ref{fig:fig1}c. 
Visually, these spots are reminiscent of labyrinthine Turing patterns, possibly caused by a reaction-diffusion system hypothesized in Fig.~\ref{fig:fig1}d. 
However, an electroconvective origin must be strongly considered, as this simple parallel-plate setup is the typical configuration used for modeling and studying electroconvection.\cite{Aleksandrov_2002,Kim2_2019,wu:hal-05540494,Luo_2018}
Therefore, this study investigates the influence of various experimental variables, including electrical signal parameters, inter-electrode distance, counter-electrode geometry, and solution viscosity, to elucidate the mechanisms governing spot organization. 
Three distinct morphologies were identified: marbled patterns, rosettes, and spots (Fig.~\ref{fig:fig1}e--g). 
Our findings suggest that electroconvection plays a predominant role in organizing the spots and determining the characteristic wavelength of the pattern, a role often underestimated in PEDOT electropolymerization. 
Finally, we explore a practical application of these spots as electrochemically deposited footprints, echoing the use of metallic dendrites for similar hardware security purposes.\cite{Kozicki_2021,kozicki2023fabrication}
When utilized as physical tags, these electropolymerized patterns may exploit the properties of conductive polymers to provide multidimensional electrochemical signatures on flexible substrates.
Ultimately, this work represents a new step toward harnessing self-organization mechanisms to fabricate low-energy-cost structures using semiconducting organic materials.
\newpage

\section{Experimental}

\subsection{Materials and methods}
All the chemicals used in this study were purchased from commercial suppliers and used as received, without any additional purification. 
They were stored and handled under standard ambient laboratory conditions (air, room temperature, and ambient light). 
3,4-Ethylenedioxythiophene (EDOT), glycerol, sodium polystyrene sulfonate (NaPSS, 70~kDa), and p-benzoquinone (BQ) were obtained from Sigma-Aldrich.

\subsection{Device Microfabrication}
Two planar electrodes were used in this study as illustrated in Figure~\ref{fig:fig1}a. 
Each consists of a 1 mm-thick 4.9~cm~$\times$~4.9~cm glass substrate. 
The bottom electrode was coated with a 100~nm gold layer deposited on a 10~nm titanium adhesion layer through e-beam evaporation using a PLASSYS MEB 550SL. 
The design of the top electrode varied depending on the experiments. 
For the study presented in Fig.~\ref{fig:fig4}, a top electrode was fabricated with a hexagonal grid using photolithography and the same metals and thicknesses. 
Starting from Fig.~\ref{fig:fig5}, the top electrode (cathode) was coated with only 12~nm of gold for 3~nm of titanium, to render it semitransparent and allow optical observation of the growth from above. 
Finally, in Fig.~\ref{fig:fig2} and Fig.~\ref{fig:fig3}, the top electrode was identical to the bottom electrode.

\subsection{Electropolymerization and Characterization}
This paper adapts to planar electrodes the AC-electropolymerization procedure used by Janzakova \textit{et al.}\cite{Janzakova2021a} on wire electrodes. 
A unipolar square wave voltage was generated by a HW-753 PWM Signal Generator Module when voltages \textit{V} above 8~V were required, and by a Modulab XM MTS (Solartron Analytical) impedance analyzer for Fig.~\ref{fig:fig6}, Fig.~\ref{fig:fig8} and Fig.~\ref{fig:fig9}. 
The growth signal was applied between the two electrodes at a given frequency \textit{f} and duty cycle dc. 
When the counter-electrode is patterned or semi-transparent, growth was observed from above the top electrode using a VGA CCD color Camera from HITACHI Kokusai Electric Inc.

\subsection{Image Capture and Processing}
The patterns presented in this work were all captured using a Keyence VHX-6000 microscope (TIFF formatted, 3$\times$8 bits RGB images). 
When detailed comparisons between images were necessary (for example in the Principal Component Analysis --- PCA --- study), the same optical conditions were used (shutter speed, working distance, no software contrast).\\ 
The PCA transformations were computed in Python using the scikit-learn library, either from the raw or all compressed images (vector dimension: \textit{k}$\times$\textit{l}$\times$\textit{l}, \textit{k}: the number of color channels, \textit{l}$\times$\textit{l}: the squared-image pixel size) or from the 9-component feature vectors \textbf{n} for each binarized image.
These feature vectors \textbf{n} = (\textit{n}\textsubscript{1}, ... , \textit{n}\textsubscript{9}) are characteristic for all binarized images of any size 1$\times$\textit{l}$\times$\textit{l}, each containing \textit{n}\textsubscript{p} pixels having \textit{p} direct neighbors of a different color.\\ 
The image compression pipeline starts with an automatic detection of the center of the pattern using a combination of blurring and thresholding to detect the pattern's outline. 
Afterwards, the pattern is cropped around its center as a square of size \textit{l}$\times$\textit{l}, ensuring that the inner regions containing patches are all contained in this square and exclude the patterns' outer ring (\textit{l} = 840). 
The length \textit{l} is then reduced to the nearest lower power of two (\textit{l} = 512), using openCV resize function and an INTER\_AREA interpolation method. 
Colors are converted to grayscale using the openCV cvtColor function. 
The thresholding method "bw\_pc" subsequently applied to binarize the pictures aims for an equirepresentation of black and white pixels in a binarized image. 
Finally, the images are successively downsampled by a factor of two (retaining one pixel every two pixels), which divides \textit{l} by two at each step. 
Each colored picture containing 3-component pixels ((r, g, b)\textsubscript{1}, ... , (r, g, b)\textsubscript{\textit{l}$\times$\textit{l}}) in a dataset were flattened (r\textsubscript{1}, g\textsubscript{1}, b\textsubscript{1}, ... , r\textsubscript{\textit{l}$\times$\textit{l}}, g\textsubscript{\textit{l}$\times$\textit{l}}, b\textsubscript{\textit{l}$\times$\textit{l}}) to build the PCA matrix. 
When considering the effect of certain alterations to the pictures (as in Fig.~\ref{fig:fig9}), the altered data were not included in the PCA matrix but were projected on the already existing space (as detailed later). 
To ensure reproducibility, a deterministic Singular Value Decomposition ---
SVD --- "full" solver was employed for the decomposition. 
Finally, the ellipses visible in the PCA results were calculated to represent the theoretical 95\% confidence region defined by the covariance for a given population; these populations were determined via k-means clustering assuming two distinct clusters.\\

\newpage

\begin{figure}[!h]
  \centering
  \includegraphics[width=1\columnwidth]{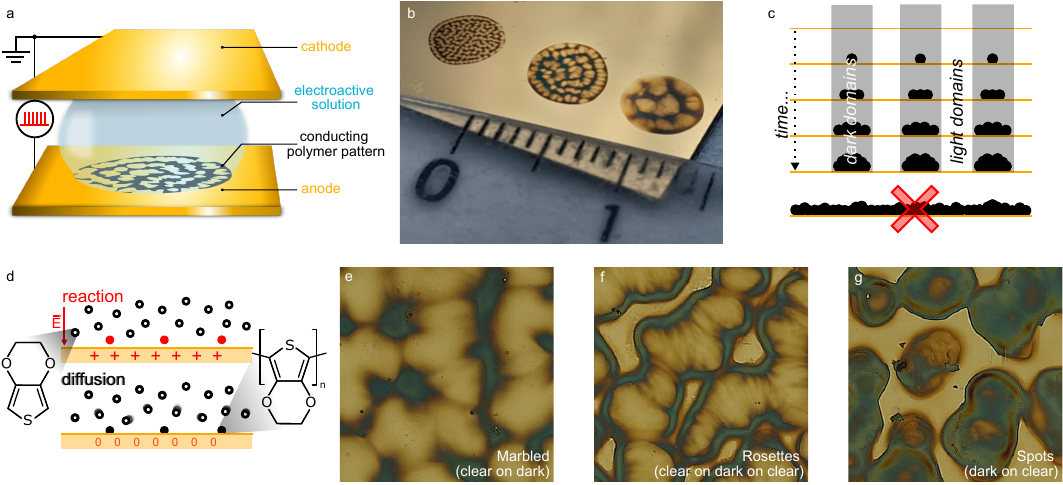}
  \caption{\textbf{Setup, mechanism and structures $\vert$}
  \textbf{a,} Experimental setup for the electrogeneration of disordered stains of conducting polymer on the bottom of a two-electrode stack, sandwiching an electro-active aqueous electrolyte containing EDOT, BQ and the NaPSS salt. 
  \textbf{b,} Photograph of three different growths electrogenerated on the same gold surface with the same electrolyte and the same voltage pattern, but different spacings between both electrode planes. 
  \textbf{c,} Unlike DC-electropolymerization, the growths performed in this study exclusively yielded electropolymerized dark domains gathering islands of uncoated gold surfaces. 
  \textbf{d,} Depiction of nucleation/growth mechanisms occurring upon unipolarization on the bottom anode, where reaction/diffusion time is conditioned by the spiking voltage pattern applied at the anode. 
  \textbf{e--g,} Microscope pictures of three classes of conducting polymer patterns, obtained with the same water-based electrolyte solution but under different physical and electrical conditions: "marbled" patterns of isolated light domains enclosed by a dark, continuous phase with diffuse boundaries (\textbf{e}), "rosettes" displaying light domains surrounded by a dark domain, which is itself surrounded by a clear continuous phase with a sharp boundary (\textbf{f}) and "spots" displaying dark domains surrounded by a clear continuous phase with a sharp boundary (\textbf{g}).
  }
  \label{fig:fig1}
  \end{figure}

\newpage

\section{Results}

\subsection{Unipolar spike voltage parameters: inherent disorder but low sensitivity}
\begin{figure}[!h]
  \centering
  \includegraphics[width=1\columnwidth]{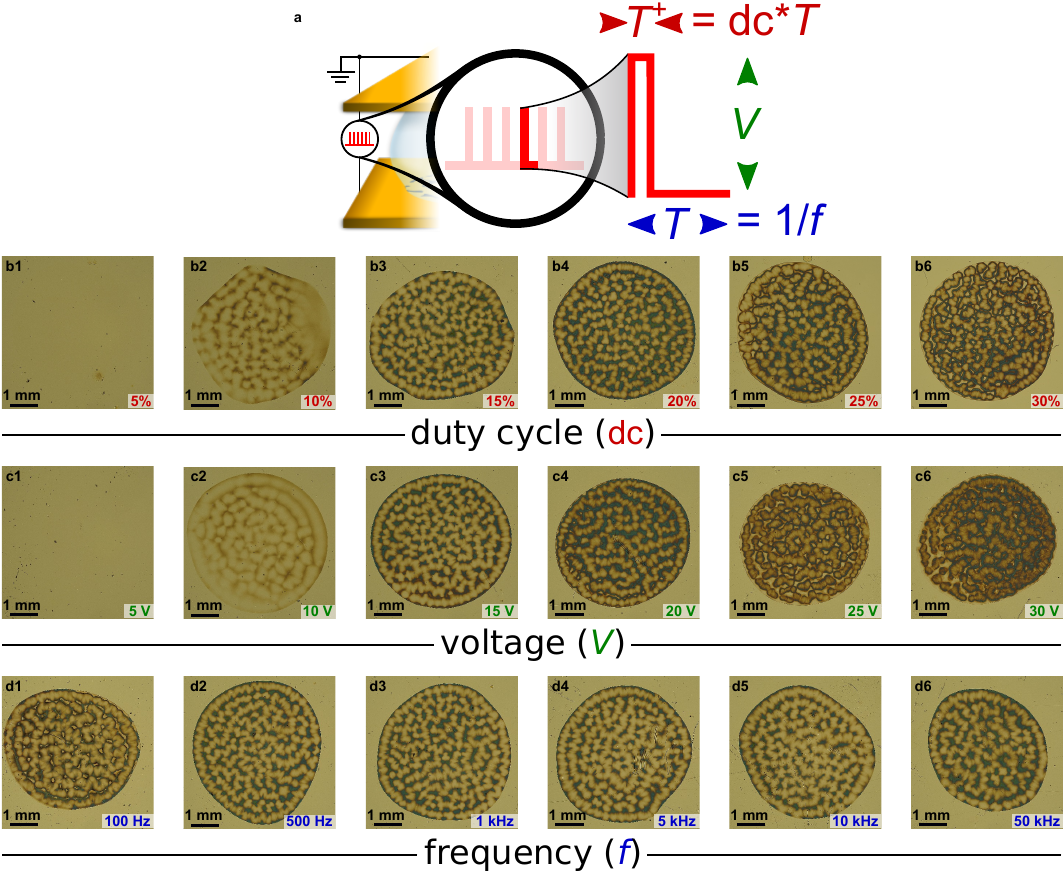}
  \caption{\textbf{Parametric study of the electrogenerated polymer patterns $\vert$}
  \textbf{a,} Pulse-width modulated waveform inputted between both substrates to electrogenerate the polymer patterns on the bottom substrate. 
  \textbf{b1--b6,} Influence of the duty cycle at 5\% (\textbf{b1}, after 1'00"), 10\% (\textbf{b2}, after 0'30"), 15\% (\textbf{b3}, after 0'15"), 20\% (\textbf{b4}, after 0'10"), 25\% (\textbf{b5}, after 0'05") and 30\% (\textbf{b6}, after 0'05"). 
  \textbf{c1--c6,} Influence of the pulse size of the voltage waveform at 5~V (\textbf{c1}, after  0'30"), 10~V (\textbf{c2}, after 0'20"), 15~V (\textbf{c3}, after 0'15"), 20~V (\textbf{c4}, after 0'10"), 25~V (\textbf{c5}, after 0'05") and 30~V (\textbf{c6}, after 0'05"). 
  \textbf{d1--d6,} Influence of the pulse frequency of the voltage waveform at 100~Hz (\textbf{d1}, after 0'15"), 500~Hz (\textbf{d2}, after 0'15"), 1~kHz (\textbf{d3}, after 0'15"), 5~kHz (\textbf{d4}, after 0'15"), 10~kHz (\textbf{d5}, after 0'15") and 50~kHz (\textbf{d6}, after 0'15"). 
  All samples were electrogenerated under ambient conditions on the same gold surface with an aqueous electrolyte composed of 10~mM of EDOT, 10~mM of BQ and 1~mM of NaPSS. 
  The standard electrode spacing was 0.3~mm, and except if mentioned otherwise, deposition was performed with a pulse waveform of 15~V, 15\% duty-cycle and a frequency of 1~kHz.
  }
  \label{fig:fig2}
  \end{figure}

To investigate whether the self-organization of PEDOT patterns -- obtained under the same conditions as PEDOT dendrites on wires -- can encode electric information, we systematically varied the parameters of the square wave signal used to generate them. 
These parameters included voltage magnitude, frequency, and duty cycle, following an approach analogous to that employed for conducting polymer dendrites.\cite{Janzakova2021a}
Fig.~\ref{fig:fig2}a\textsubscript{i} display the patterns obtained by varying the duty cycle from 5\% to 30\%. 
If no pattern is observed at 5\%, the deposit gradually darkens as the duty cycle increases, and the marbled morphology transitions to rosettes as the pulse duration increases. 
However, the size and shape of the bright areas remain constant between the experiments. 
For the highest duty cycle tested, the ring surrounding the droplet appears to be less ordered.\\

The influence of voltage magnitude on the morphology of the patterns was also investigated, as shown in Fig.~~\ref{fig:fig2}.b\textsubscript{i}. 
As previously observed when varying the duty cycle, a pattern appears only when the voltage reaches a certain threshold (10~V), and transitions from marbled patterns to rosettes as the voltage amplitude increases while keeping a relatively constant patch size between the experiments.
At high voltages, the patches at the center appear slightly bigger than those at the boundaries.\\ 
Finally, the frequency of the square-wave voltage was changed from 100~Hz to 50~kHz, without revealing obvious changes in morphology. 
Only the 100~Hz morphology is dominated by rosettes, while the rest displays marbled patterns. 
The subtle differences between the morphologies obtained when varying the parameters of the electrical signal indicate that there is a characteristic dimension of the cells around 200~µm. 
This length appears to be independent of the signal, contrary to what is observed with conducting polymer dendrites grown in similar conditions, whose thickness decreases with increasing frequency.
In a reaction-diffusion system, the spatial wavelength of the patterns, defined as the characteristic distance between adjacent repeating structures, should depend on the reaction rate and the diffusion coefficients of the chemical species,\cite{murray2003mathematical} suggesting that changes in the frequency, duty cycle, or any other temporal parameter that influences the consumption of the reactants should affect the patterns. 
In fact, reaction-diffusion systems may produce different forms, including labyrinthine or circular patterns.\cite{Turing1952} 
In the present study, applying unipolar voltage pulses over a wide range of frequencies and duty cycles did not alter the pattern wavelength nor disrupt self-organization behaviour.  
This invariance in wavelength appears inconsistent with a pure reaction-diffusion mechanism. 

\subsection{Conducting polymer patterns size with the inter-electrode gap}
\begin{figure}
  \centering
  \includegraphics[width=1\columnwidth]{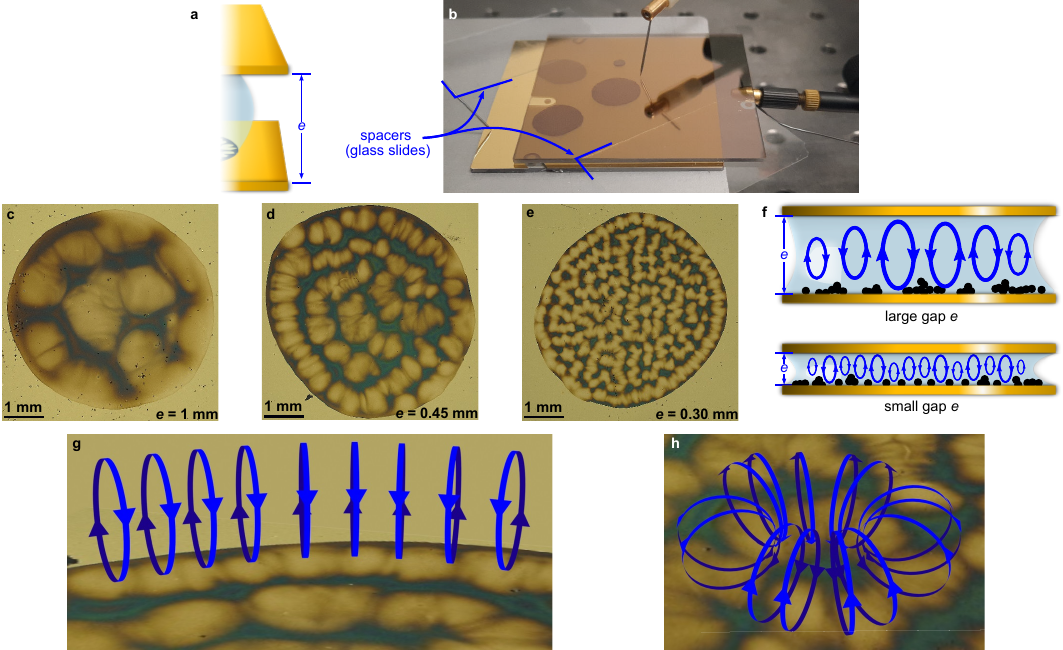}
  \caption{\textbf{Effect of the interelectrode distance on the average size of the electrogenerated patterns $\vert$}
  \textbf{a--b,} Definition of the interelectrode distance \textit{e}, controlled by the staking of spacers with specific thicknesses between both electrodes (the photograph \textbf{b} displays glass coverslips used as spacers, and a semi-transparent top-electrode). 
  \textbf{d--f,} Microscope images of three growths generated with the same electroactive electrolyte solution (10~mM EDOT, 10~mM BQ and 1~mM NaPSS in water) and the same voltage pattern (15~V, 15\% duty-cycle and 1~kHz for 0'15"), but with different interelectrode distances (\textit{e}): 1~mm (\textbf{a}), 0.45~mm (\textbf{b}) and 0.30~mm (\textbf{c}). 
  \textbf{f,} Schematics of electrohydrodynamical vortices proposed to rule the electrogenerated pattern organization, which average size directly depends on the interelectrode distance controlling the size of the vortices. 
  \textbf{g,} Hypothetical flow at the edges of the droplet showing an almost fully merged roll. 
  \textbf{h,} Hypothetical flow around the center of the droplet of partially merged vortices forming a closed-loop ring.
  } 
  \label{fig:fig3}
  \end{figure}

As shown in Fig.~\ref{fig:fig2}, the growth signal parameters (voltage, frequency and duty cycle), which strongly affect the morphology of conducting polymer dendrites grown under similar conditions, have only a minor effect on the morphology of the patterns.
This suggests a dominant formation mechanism that may differ from that of dendrites. 
The influence of the distance e between the planar electrodes (as illustrated in Fig.~\ref{fig:fig3}a) on the morphology of the deposits may provide insight into the mechanism underlying the formation of the reported patterns. 
Thus, Fig.~\ref{fig:fig3}c--e show the patches' morphologies obtained for three electrode gaps: 1~mm, 0.45~mm, and 0.3~mm. These gaps are realized using glass spacers as pictured in Fig.~\ref{fig:fig3}b.
It appears immediately that the larger the gap, the larger the resulting patches. 
Electroconvection has already been proposed as a mechanism explaining the formation of spots during electrodeposition.\cite{Pashchanka_2021} 
An electroconvective origin is plausible, first because of the high voltages involved ($\geq$ 10~V), since the space-charge region responsible for electroconvective instabilities is favored at increased voltages,\cite{ZALTZMAN_2007} and second because the experimental setup resembles the configuration most commonly used to model electroconvection (parallel planar electrodes).\cite{Luo_2018,Han_2006}
The patterns themselves resemble those obtained in both simulations,\cite{Demekhin_2014} and experiments.\cite{Stockmeier_2023}
The observed dependence on the gap could be explained by the restriction of the expansion of electroconvective vortices.\cite{Davidson_Mani_2017}
Indeed, the patch size remains on the same order of magnitude as the gap (Fig.~\ref{fig:fig3}c-–e). 
The electroconvection length scale is comparable to the diffusion layer that develops by depletion of a reactant at the interface.\cite{Yossifon_2008}
The upper electrode may thus act as a barrier to the expansion of the diffusion layer, reducing the size of the patches. 
Although vortices in electroconvection have been reported to increase in size with the applied voltage in certain systems,\cite{Rubinstein2008} we do not observe such a trend with the patterns obtained in our setup. 
This does not necessarily contradict an electroconvective origin, as the characteristic size may be primarily set by the electrode gap, which appears to limit the maximum achievable size in all conditions shown in Fig.~\ref{fig:fig2}b–-d. 
The possibility of producing patterns at very small scales therefore becomes conceivable by controlling the electrode gap. 
Fig.~\ref{fig:fig3}f illustrates how the presence of vortices could affect the distribution of species in solution (monomer, ions and particles, charged or neutral). 
The fluid, stirred by loops, could carry only a minimal quantity of species, while most of them may accumulate between the loops. 
As the electropolymerization reaction proceeds, particles of increasing sizes may be produced, which may tend to exclude them from the core of the vortices due to fluid inertia.\cite{Aulnette_2025}\\
In the context of electroconvection, the peripheral ring of patches in the droplet may result from the coalescence of individual vortices into a continuous convective roll, as depicted in Fig.~\ref{fig:fig3}g. 
Indeed, the presence of a lateral boundary can influence the orientation of convective structures within the fluid, often forcing a pattern that reflects the shape of the boundary.\cite{Shishkina2021}
Within the central region of the droplet, structures similar to rolls are seen at certain locations (Fig.~\ref{fig:fig3}h), sometimes forming closed-loop rings. 
It is hypothesized that these patterns result from the spatial packing of confined vortices.

\newpage

\subsection{Order level controlled with a periodical counter-electrode}
In the previous sections, electropolymerization was conducted with identical counter- and working electrodes with no means of monitoring the reaction from above. 
Consequently, it was impossible to directly assess whether gas bubbles contributed to the observed patterns. 
However, their presence is highly probable since the high voltages involved go beyond the limited electrochemical window of water. 
In this section, the counter-electrode (CE) was patterned with a hexagonal mesh providing optical transparency while maintaining electrical conductivity across the surface.

\begin{figure}[!h]
  \centering
  \includegraphics[width=1\columnwidth]{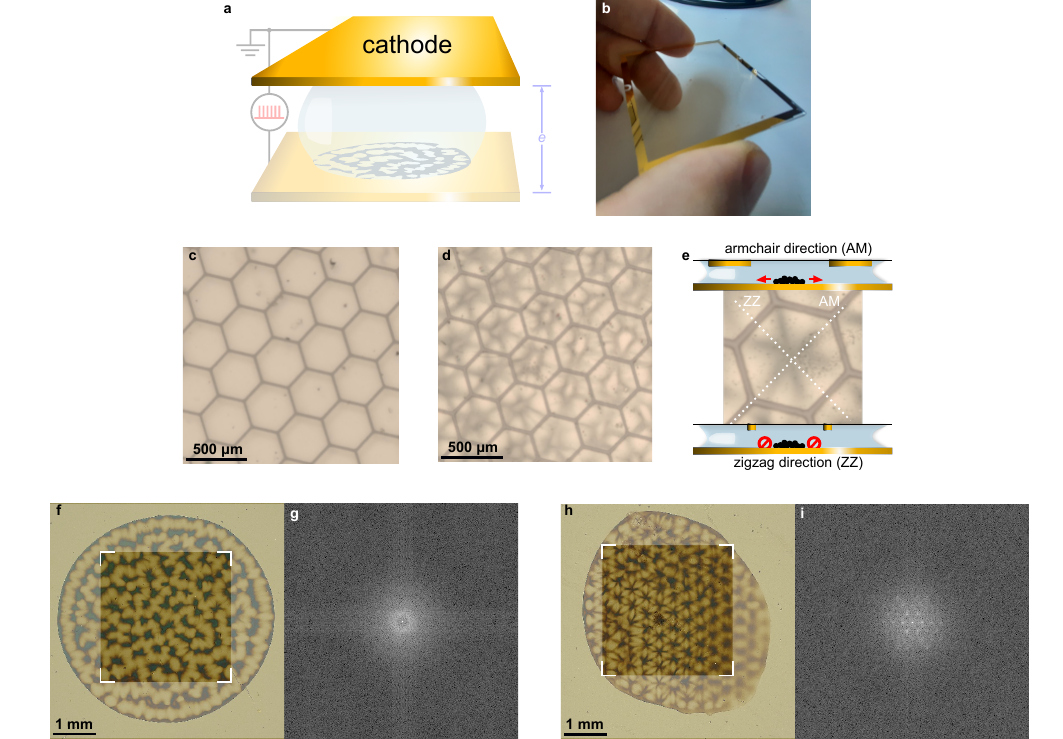}
  \caption{\textbf{Controlled disorder with a periodically structured cathode $\vert$}
  \textbf{a--b,} Impact of the cathode (top-electrode) on the morphological control of the electrogenerated patterns at the bottom electrode (in \textbf{b}, a photograph of the honeycomb-structured gold cathode used in the following experiment). 
  \textbf{c,} Microscope picture of the electrochemical system containing the honeycomb-structured cathode (periodicity: 1/366~$\upmu$m\textsuperscript{-1}). 
  \textbf{d,} Microscope picture of the electrochemical system as seen through the CE after electrogeneration. 
  \textbf{e,} Schematics of the star-like growth electrogenerated from the honeycomb-cathode, with orientation and periodicity matching the lattice of the cathode. 
  \textbf{f,} Microscope picture of a disordered pattern grown on a fully-coated gold anode and with a fully-coated gold cathode. 
  \textbf{g,} Fast Fourier Transform of the disordered pattern, computed from the square zone limited by the white vector frame corners in \textbf{f}. 
  \textbf{h,} Microscope picture of a periodical pattern grown on a fully-coated gold anode and with a honeycomb-structured gold cathode. 
  \textbf{i,} Fast Fourier Transform of the periodic pattern, computed from the square zone limited by the red vector frame corners in \textbf{h}. 
  Both cases illustrated in \textbf{f--i} were generated with an electrode spacing was 0.3~mm, stopped after 0'15" and performed with a pulse waveform of 15~V, 15\% duty-cycle and a frequency of 1~kHz. 
  Both experiments were performed with the same electroactive electrolyte solution (10~mM EDOT, 10~mM BQ and 1~mM NaPSS in water).
  } 
  \label{fig:fig4}
  \end{figure} 

The electrode is shown at macroscopic scale in Fig.~\ref{fig:fig4}b, and under experimental conditions above the droplet in Fig.~\ref{fig:fig4}c, confirming its transparency. 
Since the electrode is not homogeneous, it may concentrate the electric field in certain areas which could lead to dendritic growth or affect the electroconvective flow. 
Fig.~\ref{fig:fig4}d shows a picture of the deposit, revealing that the hexagonal mesh influenced the deposition by concentrating the PEDOT directly below the center of the hexagons with a marbled morphology, as visible in Fig.~\ref{fig:fig4}d. 
Meanwhile, when a rosette morphology is obtained, the center appears clearer (Fig.~\ref{fig:fig4}d). 
This suggests that the vortices are oriented mainly toward the conducting areas of the counter-electrode, while PEDOT particles may deposit preferentially below the glass, non conducting areas. 
PEDOT may also accumulate along the diagonals going through the vertices of the hexagon (Fig.~\ref{fig:fig4}e). 
It is hypothesized that these directions correspond to boundaries between adjacent aligned vortices, which are also areas where PEDOT may accumulate. 
The selected wavelength appears to be dictated by the CE patterning. 
This length scale is of the same order of magnitude as the wavelength observed in the unpatterned configuration, probably slightly smaller, as indicated by the narrower dark regions separating the cells.
Although gas bubbles may eventually be observed, they only appear around 15 seconds after the first pulse is applied, while the pattern forms within the first five seconds. 
This time lag suggests that bubbles do not cause the observed patterns. 
When bubbles exist before applying a voltage (for example, as air pockets trapped when the counter-electrode is placed on top), they act as physical boundaries around which the convective cells organize.\\
A two-dimensional Fourier transform was applied to the pattern obtained using a uniform counter-electrode (Fig.~\ref{fig:fig4}f) and the one obtained using a hexagonal counter-electrode (Fig.~\ref{fig:fig4}h). 
The 2D spectra are presented in Fig.~\ref{fig:fig4}g and Fig.~\ref{fig:fig4}i, respectively. 
For the uniform electrode, a ring dominates the spectrum around the center, which indicates that the deposition does not have a preferential direction. 
The broadness of the ring (200~$\upmu$m) suggests that the spacing between the bright patches varies significantly across the pattern. 
In contrast, when using the electrode with a hexagonal mesh, the frequency content concentrates into discrete peaks distributed along six principal directions separated by an angle of $\pi$/3. 
These directions feature harmonic and fundamental peaks. 
The fundamental component corresponds to a spatial period of around 318~$\upmu$m, which matches the distance between each dark patch as well as the characteristic periodicity of the upper electrode's hexagonal mesh (320~$\upmu$m). 
Thus, the upper electrode seems, in this case, to impose its periodicity on the pattern. 
These experiments show that electroconvection is also present in this configuration but that the flow distribution is clearly affected by the patterning of the CE. 
The cell distribution, stochastic in the case of an unpatterned CE, becomes more ordered with a periodic CE.\\

\subsection{Working electrode roughness and direct implementation on a circuit board}
Since the electropolymerized patterns are conductive, the information they encode can be accessed electrically. 
If such a pattern was deposited on a grid of electrodes individually addressable using vias,\cite{Perdigones_2022} and if the pixel density was high enough, a discrete pattern could be electrically readable. 
In this perspective, printed circuit board (PCB) technology appears to be particularly suitable for low-cost fabrication of such grids without requiring cleanroom or resource-intensive microfabrication processes.\cite{Mohd_Asri_2021} 
Until this point, patterns were always obtained using e-beam evaporated gold offering a well-controlled gold coating. 

\begin{figure}[!h]
  \centering  \includegraphics[width=1\columnwidth]{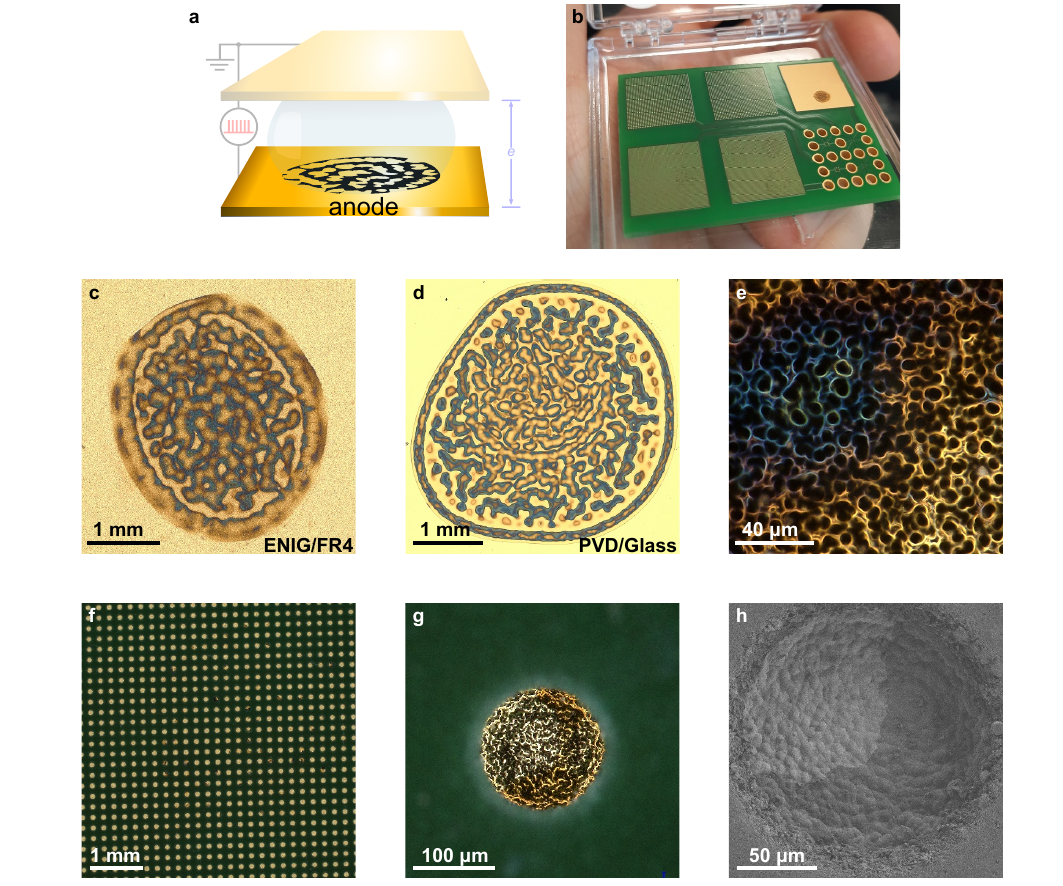}
\caption{\textbf{Electrochemical growth on a rough surface $\vert$}
\textbf{a--b,} Impact of the anode (bottom electrode) on the morphological control of the electrogenerated patterns (in \textbf{b}, photograph of a commercial ENIG coated copper electrodes on FR4, where the electrochemical growths were performed). 
\textbf{c--d,} Microscope images of two electrochemical growths performed under the same conditions of voltage and concentrations (\textit{V}~=~5~V, 50\% duty cycle, \textit{f}~=~1~kHz, 1~mM NaPSS, 10~mM EDOT/BQ in water), on ENIG coated copper electrodes on FR-4 (\textbf{c} - rough electrode) or evaporated gold on a glass substrate (\textbf{d} - flat electrode). 
\textbf{e,} Microscope image of an ENIG-coated electrode after electropolymerization, at the boundary between a dark and a light area. 
\textbf{f,} Microscope image of remaining dark stains after an attempt to electrogenerate polymer growth on a array of ENIG plated contacts. 
\textbf{g,} Zoom-in on one contact where deposition was observed. 
\textbf{h,} Scanning Electron Microscope image of the same contact.
}
  \label{fig:fig5}
  \end{figure} 

Meanwhile, PCBs are typically finished with an electroless nickel immersion gold (ENIG) layer, a more accessible process that results in a less controlled roughness.\cite{waseem2023enig} 
Considering the high voltages involved in pattern growth, nickel that is imperfectly covered by the gold layer could oxidize and grow nickel dendrites.\cite{Singh2020}
Moreover, high roughness may lead to conducting polymer dendrites if the distinction between CPDs and patterns originates from local defects. 
To test these hypotheses, the same growth voltage that had previously yielded patterns was applied to a PCB as shown in Fig.~\ref{fig:fig5}b, which includes a full ENIG-coated face and an array of circular holes with a diameter of approximately 150~$\upmu$m. 
These electrodes were used as the working electrode with a semi-transparent counter-electrode. 
An aqueous solution containing 10~mM EDOT and 1~mM NaPSS was used. The pattern printed on the full ENIG-coated surface is presented in Fig.~\ref{fig:fig5}c, next to the pattern grown under the same conditions on e-beam evaporated gold (Fig.~\ref{fig:fig5}d). 
The cell arrangement is slightly different on ENIG, with larger cells more tightly packed together. 
A magnified view of the surface is presented in Fig.~\ref{fig:fig5}e. 
The film exhibits good contrast and appears to be unaffected by the roughness of the ENIG surface. 
A test conducted in water on an array of holes led to visible deposition in some of the holes (Fig.~\ref{fig:fig5}f) but the deposited film inside appears discontinuous (Fig.~\ref{fig:fig5}g). 
It may have been damaged by gas bubbles. 
An SEM image (Fig.~\ref{fig:fig5}h) reveals a well approximately 35~$\upmu$m deep with solder-mask walls that are not perfectly vertical.\\
Is it hypothesized that to grow patterns on an array of holes, either their surface needs to be planar or specific geometric conditions on the diameter and spacing of the holes may be required. 
In this case, the diameter and spacing are both comparable in length to the patch size, which may affect the distribution of vortices and impact how PEDOT:PSS will deposit inside the holes. 
Pattern growth is therefore possible directly on ENIG but may require specific size adaptations.\\

\subsection{Influence of the chemical composition of the solution on the conducting polymer patterns}
\begin{figure}
  \centering  \includegraphics[width=1\columnwidth]{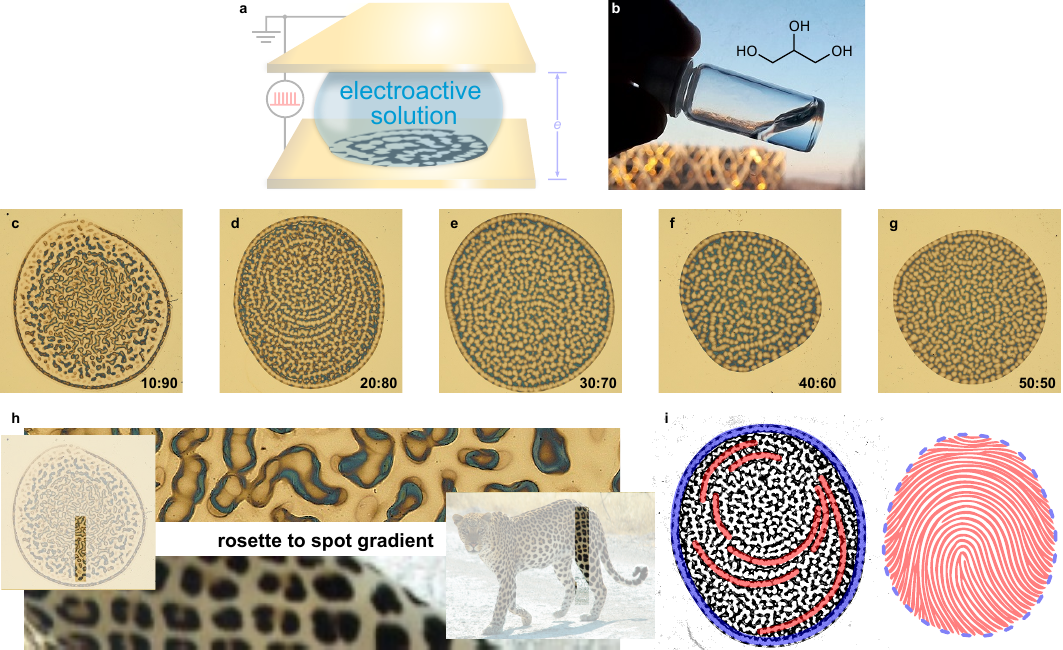}
  \caption{\textbf{Dependence of morphological patterns on the chemical composition of an electroactive aqueous electrolyte solution. $\vert$ }
  \textbf{a--b,} Impact of the electroactive solution on the morphology of the electrogenerated patterns (in \textbf{b}, photograph of a vial of glycerol used as a water-miscible viscous cosolvent, in which the electrochemical growths were performed).
  \textbf{c--g,} Microscope images of five electrogenerated patterns grown in the same electrical conditions (\textit{V}~=~5~V, 50\% duty cycle, \textit{f}~=~1~kHz), with electro-active electrolyte solutions (10~mM EDOT, 10~mM BQ and 1~mM NaPSS) prepared with different volume ratio of glycerol:water as a solvent mixture: 10:90 (\textbf{c}), 20:80 (\textbf{d}), 30:70 (\textbf{e}), 40:60 (\textbf{f}) and 50:50 (\textbf{g}).
  \textbf{h,} Zoom-in on Fig.~\ref{fig:fig6}c (with 10\% glycerol) showing the pattern contrast between the inner area and the outer area on identical coatings, which resembles the ones on animal furs.\cite{leopard}
  \textbf{i,} Contrasted black-and-white image of Fig.~\ref{fig:fig6}d (with 20\% glycerol) stressing on the concentric organization of some electrogenerated patterns, which resembles the ones observed in fingerprints. Fingerprints picture: Gordon Johnson / pixabay.\cite{fingerprint}
  }
  \label{fig:fig6}
  \end{figure}

If electroconvection is the dominant mechanism behind the formation of these patterns, it is expected that the morphology of the patches will be influenced by parameters such as the viscosity or the density of the solvent, on which electroconvective instabilities depend.\cite{Aleksandrov_2002}
In particular, viscosity should decrease the velocity of the vortices,\cite{Grinbank_2009,Sasmal_2022} which may affect the shape of the cells. 
The viscosity of the previously used aqueous solution can be increased by adding glycerol (as shown in Fig.~\ref{fig:fig6}b), leading to different patterns under identical voltage and gap conditions as seen in Fig.~\ref{fig:fig6}c--g for five different glycerol-to-water ratios. 
The morphology seems to evolve mainly between 10\% and 30\% glycerol. 
At 10\%, rosettes dominate, and the pattern transitions to marbled structures as more glycerol is added.

In addition, the pattern at 10\% glycerol shows a contrast in morphologies between the center, dominated by rosettes, and the periphery, dominated by bright patches. 
This type of contrast is reminiscent of how spots on a leopard's fur vary, as illustrated in Fig.~\ref{fig:fig6}h, where rosettes become black spots around the legs and belly. 
Although evaporation may lead to a radial gradient within the droplet, the setup minimizes the surface of the droplet in contact with air by confining it, making it unlikely for the observed contrast to be caused by evaporation. 
The ring surrounding the droplet exhibits a roll-like structure (like a striated ring) at 20\% and 30\% glycerol content, transitioning to a pattern of distinct and well-separated cells above 30\% glycerol content. 
Increasing the glycerol fraction homogenizes the pattern as the ring and boundary with the inner cells disappear. 
This suggests that the formation of convective rolls at the edges of the droplet is suppressed or at least delayed when adding glycerol to the solution. 
The same morphological transition from marbled structures to rosettes was observed when increasing the voltage or the duration of a voltage pulse (Fig.~\ref{fig:fig2}). 
These parameters, such as low viscosity, promote the formation of convective structures such as  vortices or rolls, as well as the deposition of PEDOT:PSS on the electrode. 
Thus, it can be hypothesized that the emergence of either marbled or rosette morphologies is decided by an interplay between vortex dynamics (expansion and velocity) and the electropolymerization reaction. 
Interestingly, some of the inner cells in all patterns above 10\% glycerol organize into ordered structures similar to curved rolls. 
This, in turn, is reminiscent of human fingerprints, as illustrated in Fig.~\ref{fig:fig6}i. 
Those are unique to each individual, offering a possible means to uniquely identify a given pattern. 

\newpage

\subsection{From conducting polymer patterns to fingerprints}
\begin{figure}[!h]
  \centering
  \includegraphics[width=1\columnwidth]{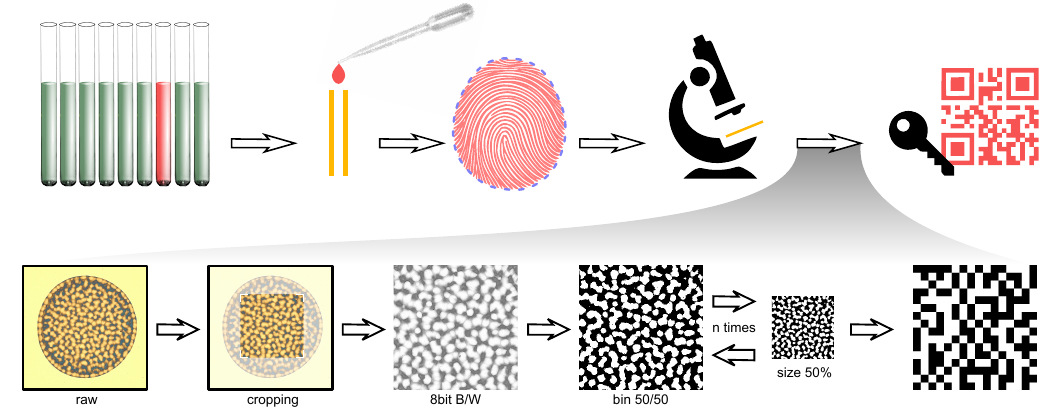}
  \caption{\textbf{Exploiting electropolymerized fingerprints to engrave the information of a solution composition $\vert$ } 
  As conducting polymer patterns are unique and specific to the chemical composition of a solution, the following study proposes to test their potential as fingerprints to classify them by chemical composition on two levels (representative of a lab test to identify true positive and true negative liquid samples). Fingerprints picture: Gordon Johnson / pixabay.\cite{fingerprint} 
  Moreover, we investigate on how deep can these fingerprint images be compressed while preserving the chemical character of the imprinted solution (in the subsequent Fig.~\ref{fig:fig8}) and study how the information on these fingerprints is resilient to optical and physical changes (in the subsequent Fig.~\ref{fig:fig9}).
  } 
  \label{fig:fig7}
\end{figure}

Previously, we highlighted that PEDOT:PSS, when electrosynthesized with spike voltages between two plates, forms peculiar patterns (Fig.~\ref{fig:fig1}). 
The formation of these patterns is weakly influenced by voltage conditions (Fig.~\ref{fig:fig2}) and substrate roughness (Fig.~\ref{fig:fig5}).
As long as the distance between both plates remains steady (Fig.~\ref{fig:fig3}) and the counter-electrode is uniform (Fig.~\ref{fig:fig4}), the imprinted pattern shows well-defined randomly-organized morphologies.
When the composition of the electrolytic solution changes, the physical boundaries of the process are altered, causing different patterns to be imprinted on the anode surface (Fig.~\ref{fig:fig6}).
Given the low energy requirements and the ease of implementing electropolymerization without specialized equipment, these patterns can be used to produce unique identifiers. 
The chemical identity of a solution could potentially be encoded in the taxonomy of various pattern families, their gradients, or local organizations, all of which could be engraved under well-defined physical conditions (Fig.~\ref{fig:fig7}).

\begin{figure}[!h]
  \centering
  \includegraphics[width=1\columnwidth]{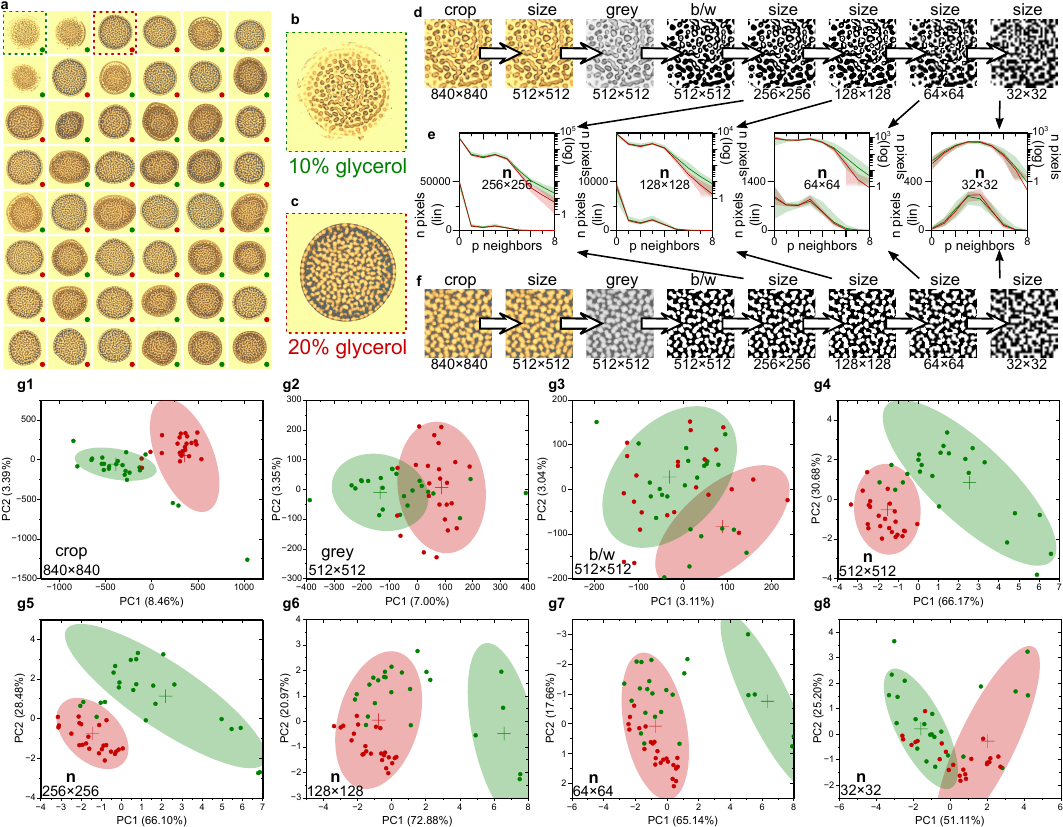}
  \caption{\textbf{Information compression and concentration classification with electropolymerized fingerprints $\vert$ }
  \textbf{a,} 48 microscope pictures of conducting polymer patterns on a common gold substrate, electrogenerated (\textit{V}~=~5~V, 50\% duty cycle, \textit{f}~=~1~kHz) with a solution (1~mM NaPSS, 10~mM EDOT, 10~mM BQ), containing either 10\% (marked with a green dot) or 20\% (marked with a red dot) of glycerol as a co-solvent in water.
  \textbf{b--c,} As illustrative examples, microscope pictures (raw) of the first patterns in the series, which are electrogenerated with either 10\% (\textbf{b}) or 20\% (\textbf{c}) glycerol solutions.
  \textbf{d,f,} Images of both illustrative examples (\textbf{d} for the 10\% example from the raw picture displayed in \textbf{b} and \textbf{f} for the 20\% example from the raw picture displayed in \textbf{c}) at different stages in the compression process.
  \textbf{e,} Pixel counting for the different binarized images at different image sizes \textit{l}$\times$\textit{l}, for the total number of pixels \textit{n} that are surrounded by \textit{p} of their eight closest neighbors that have a color different from their own. 
  The curves represents the statistic (solid line for mean, semitransparent volume for min/max range) for the populations of 48 compressed images for 10\% glycerol solutions (in green) and for 20\% glycerol solutions (in red).
  The set of curves on top are in logarithmic scale of \textit{n}, the one at the bottom are in linear scale of \textit{n}.
  \textbf{g,} PCA scores projected on the first two principal components for different populations of 48 compressed images, showing the separability of the two populations of 10\% glycerol (in green dots) and 20\% glycerol (in red dots) conducting polymer patterns (The red and green ellipses correspond to 95\%-confidence ellipses determined with k-means clustering using \textit{k}~=~2).
  } 
  \label{fig:fig8}
\end{figure}

In this section, we consider two growth solutions of different viscosities: one with 10\% glycerol and one with 20\%. 24 electrodepositions are conducted in each of the two solutions to determine the dispersion of the patterns generated from the same chemical environment, and whether an unsupervised algorithm can extract by itself sufficiently distinct optical features to trace a given pattern back to its original growth medium. 
The result could have practical implications on the potential of this method for applications such as object identification and hardware security strategies.
The two populations of 24 patterns were generated while ensuring that the experimental conditions remained under control and that bias was minimal.
The same voltage, interelectrode gap, substrate and counter electrode were used throughout the growth experiments.
While conducting electropolymerization in parallel would have been ideal, a sequential approach was adopted for the 48 droplets to avoid experimental variability caused by the delay between droplet deposition and the start of the growth process. 
Indeed, the time delay between deposition and voltage polarization was found to have an effect, presumably due to solute evaporation.
Therefore, to maintain the highest control while minimizing acquisition time, 3+3 different coatings using the two glycerol fractions were electropolymerized simultaneously on the same substrate. 
This process was replicated eight times on the same substrate.
The deposition order was alternated between runs to decorrelate the time-delay bias from that of the chemical identity of the six solutions.
Despite visualizing substantial differences between patterns belonging to the same class (the 48 images are provided in a lossless format as supplementary information in Fig.~\ref{fig:figS1}), a good information descriptor for classification seems to be one which successfully discriminates rosettes (characteristic of the 10\% population) from marbled patterns (characteristic of the 20\% population).\\
Image classification was attempted based on the composition of the pixels, particularly using the feature vector \textbf{n} where each component represents the frequency of pixels having a specific number of adjacent neighbors with an opposite polarity (black or white).
Reduced down to 9-dimensional vectors (Fig.~\ref{fig:fig8}e), the \textit{n}\textsubscript{i} coordinates of each vector characterizing a pattern image was evaluated after compressing the images at different levels (Fig.~\ref{fig:fig8}d,f): 
First, the colored image is cropped at the centre of the pattern as a square of \textit{l}$\times$\textit{l} pixels (\textit{l}~=~840, number of possible images (256\textsuperscript{3})\textsuperscript{840$\times$840}).
Then, the image is downsized to the largest power of two (\textit{l}~=~512, number of possible images (256\textsuperscript{3})\textsuperscript{512$\times$512}).
Afterward, the image is converted to grayscale (number of possible images 256\textsuperscript{512$\times$512}), and binarized (number of possible images 2\textsuperscript{512$\times$512}).
The image is subsequently downsized to \textit{l}~=~256, 128, 64 and 32 (number of possible images 2\textsuperscript{\textit{l}$\times$\textit{l}}).
Fig.~\ref{fig:fig8}d and Fig.~\ref{fig:fig8}f show the gradual degradation of the information on the pictures with the compression for two images representative of both classes.
It is expected that iteratively removing information on the color distribution of the coatings or on the contours of each pattern creates significant hindrances for a classifier attempting to distinguish between the two classes, and that the ability to identify the nature of a solution is limited by the resolution of the tool capturing an image of the pattern.
Also, in Fig.~\ref{fig:fig8}e, one can observe that the \textit{n}\textsubscript{i}\textsuperscript{th} distribution of \textit{p} neighbors is significantly affected by the size reduction of a binarized image.
Reducing the image resolution down to the characteristic size of the imprinted pattern imposes therefore a physical limit to recognize both populations under that criterion.\\
To classify the images, PCA was used as a linear classifier (see Fig.~\ref{fig:fig8}g).
PCA was performed for different compressions in order to evaluate under which criteria an image starts to lose the information which is characteristic of the chemical nature of the solution that has generated the patterns.
After cropping, this information is embedded in the first two principal components (Fig.~\ref{fig:fig8}g1).
Converting the color image to grayscale causes both populations to overlap in the PC1-PC2 score plot, which indicates that the color stores part of the information characterizing both populations (Fig.~\ref{fig:fig8}g2).
Binarization of gray pixels causes serious information loss in the projection of scores on the first two principal components. 
This is consistent with the qualitative assessment that the two populations become harder to distinguish by the naked eye after black and white conversion (Fig.~\ref{fig:fig8}g3).
Now, instead of directly using the value of a pixel to define a vector coordinate, we convert each image to the reduced 9-dimensional feature vector \textbf{n}.
For any size \textit{l}, the transformation from the binarized image to its feature vector causes a drastic diminution of vector possibilities --- from 2\textsuperscript{\textit{l}$\times$\textit{l}} dropping down to a number lower than $\binom{l^2 + 8}{8}$.
Therefore, even in the case of \textit{l}~=~32, the information compression is tremendous --- from a set of over 10\textsuperscript{308} possibilities to less than 10\textsuperscript{20}.

However, this transformation improves pattern classification, as shown in Fig.~\ref{fig:fig8}g3 and Fig.~\ref{fig:fig8}g4, despite reducing the set.
Both populations become separable with the same classifier, which testifies that the information characteristic of the local surrounding of electrodeposited polymer patterns can be used as a feature to build solution fingerprints, characteristic of their chemical composition despite their variability.
Further reduction in image resolution prior to extracting the feature vector \textbf{n} causes the information to disappear again, as image resolution starts to negatively impact the contour of the spot and marbled patterns (see Fig.~\ref{fig:fig8}g5--g8). 
While the 95\%-confidence ellipses do not overlap if \textit{l}~=~512 or 256, for lower resolution images both classes start to lose separability on the PC1-PC2 score plane: the red and green ellipses, determined by k-means clustering, poorly match the actual populations as the data are compressed.
These results reveal a trend caused by compression on classification rather than a fundamental limit for information separability in these electropolymerized patterns:
By the naked eye, one can still identify black-and-white images, even when compressed down to \textit{l}~=~128 or lower.
Neither the \textbf{n} feature vector nor the PCA architecture are optimized for this application, and more advanced supervised classifiers would certainly improve furthermore pattern identification.

\begin{figure}[!h]
  \centering
  \includegraphics[width=1\columnwidth]{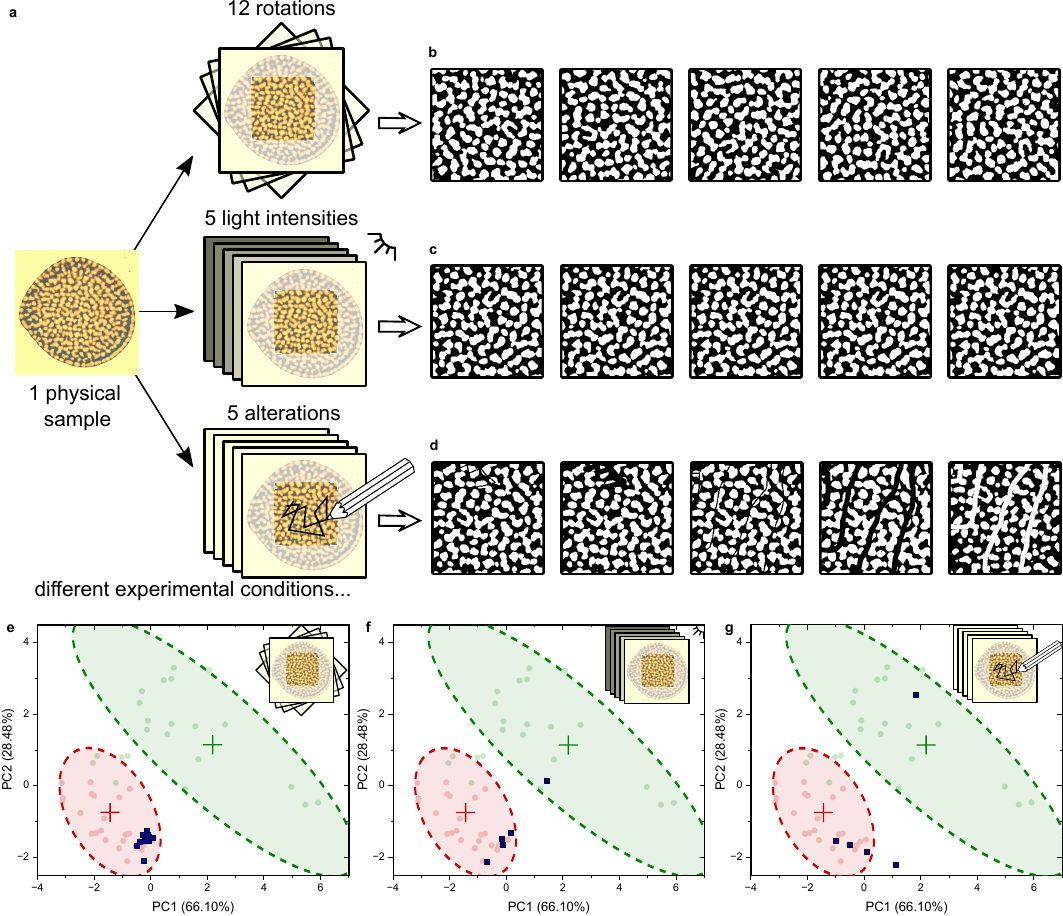}
  \caption{\textbf{Recognizing an electropolymerized fingerprint under different optical and physical perturbations $\vert$ }
  \textbf{a,} Three different cases were studied to evaluate the persistence of the chemical identity on a fingerprint, when conducting polymer patterns are photographed in different conditions:
  (1) 12 microscope pictures of a single conducting polymer pattern (previously studied in Fig.~\ref{fig:fig8} and belonging to the 20\% class) have been recorded at various azimuthal angles (the experimental conditions were kept as consistent as possible with those used for the original image in Fig.~\ref{fig:fig8}).
  (2) Five microscope pictures of the same conducting polymer pattern, generated with different light intensities under the microscope (the experimental conditions were kept as consistent as possible with those used for the original image in Fig.~\ref{fig:fig8}).
  (3) The original microscope pictures of this same conducting polymer pattern were numerically corrupted by five different pixel alterations on the raw image. 
  \textbf{b--d,} Sets of five compressed images (binarized, \textit{l}~=~256) used as fingerprints for the chosen conducting polymer pattern (raw image displayed in Fig.~\ref{fig:fig9}a) to illustrate the impact of the different physical and optical conditions (rotation in \textbf{b} and light intensity in \textbf{c}) and alteration (in \textbf{d}) on the fingerprint.
  \textbf{e--g,} PCA scores projected on the first two principal components for different populations of 48 compressed images illustrated in Fig.~\ref{fig:fig8} (binarized, \textit{l}~=~256). From the first two principal component loadings, the blue scatter dots illustrate the coordinates of the new fingerprints, taken under different physical and optical conditions (rotation in \textbf{e}, light intensity in \textbf{f}) and alteration (in \textbf{g}).
  } 
  \label{fig:fig9}
\end{figure}

To assess the possibility for the electrogenerated patterns to keep their identity under various conditions, one raw example out of the previous set of 48 patterns was intentionally biased or corrupted (Fig.~\ref{fig:fig9}).
Three different kinds of alterations were made on the electropolymerized pattern (Fig.~\ref{fig:fig9}a), displayed in the supplementary Fig.~\ref{fig:figS1}f3, belonging to the "20\% glycerol" class. 
A new subset of 12 microscope images was recorded, with different rotations of the substrate (Fig.~\ref{fig:fig9}a,b). 
Another subset of five microscope images was recorded under different light intensities (Fig.~\ref{fig:fig9}a,c). 
A third subset of images was recorded from the original microscope raw image, after digital degradation with a black or a white pen to emulate physical scratches of various lengths, thicknesses and structure (Fig.~\ref{fig:fig9}a,d).
The resulting projection on the former PCA score plot is shown in Fig.~\ref{fig:fig9}e--g, revealing the impact of the different alterations/corruptions on the identity of the pattern.\\
In general, rotating the substrate does not seem to have a significant effect on the identity of the fingerprint (Fig.~\ref{fig:fig9}e), as the newly generated scores are projected in a very confined area within the ellipse of the 20\% cluster.
As the pixel composition is significantly different in all images (Fig.~\ref{fig:fig9}b), this stresses the fact that the identity of the pattern is not engraved in the positioning of particular black and white pixels over the total area of a fingerprint.
We note a shift compared to the initial coordinates that we attributed to experimental variables other than the substrate orientation.\\
When the pattern is exposed under the same microscope but with different light intensities, the identity of the pattern is significantly affected (Fig.~\ref{fig:fig9}f), despite the fact that the images remain similar to the naked eye (Fig.~\ref{fig:fig9}c).
Although the PCA performed on \textbf{n} is not affected by the substrate orientation, this classifier may be unsuitable for images acquired at different times under varying ambient light conditions.\\
Finally, scratching the surface shows a mitigated impact on the classification (Fig.~\ref{fig:fig9}g).
Although the defects are local (Fig.~\ref{fig:fig9}d), they have different effects on the contours of the patterns at binarization.
As a consequence, two cases out of five fall outside the "20\% glycerol" class ellipse, while three of them remain contained in it.
This result shows that even if a pattern is degraded over time due to repeated handling, it may not be an ultimate limitation to retrieve the identity of an altered fingerprint.

\section{Conclusions}
Dynamic electropolymerization emerges as a viable approach to mimic natural growth mechanisms. 
It can either generate three-dimensional conducting polymer dendrites on anisotropic electrodes, or two-dimensional, self-organized conducting polymer patterns on planar electrodes using the same conditions of voltage and chemical composition. 
The resulting planar patterns closely resemble animal coat patterns. 
In this study, they were evidenced to be influenced by electroconvection, with a variety of morphologies related to the vortices that take place in solution. 
In particular, a substantial contrast exists between the ring, with ordered, parallel cells, and the inside of the droplet, where cells are organized in a more stochastic way. 
This stochasticity can serve to project information related to the chemical composition of an electroactive solution on a vast information space, which could allow the use of these patterns as fingerprints for identification, encrypted in the imprinted patterns. 
Without prior knowledge of how the patterns were generated, it was demonstrated that the class of solutions that was used to produce the patterns could be identified.\\
These patterns can be produced with low energy and chemical resources at large scale and by any end-user to create a unique personal ID. 
Similarly to electrodeposited dendrites with which they share common growth mechanisms, these patterns could be employed as an anti-counterfeiting stamp to guarantee the authenticity of goods.\cite{Kozicki_2021,kozicki2023fabrication}
Although the deposition method requires a conductive substrate, these conducting polymer patterns may in theory offer multiple advantages, such as robustness (a threefold optical, electrical and chemical contrast) and high integrability for flexible and biocompatible applications. 
From optical tags,\cite{kozicki2023fabrication} to physically unclonable functions (PUFs),\cite{Kim2022,Kayaci2022,CarroTemboury2018,Zhong2022,Dodda2021,John2021,Gao2022,Scholz2020,Li2022,Wali2019,Shamsoshoara2020,Gebali2022,Arppe2017,Maes2010} to information storage on (bio)polymers,\cite{CRCHIM_2021__24_1_69_0,Church_2012} this new class of conducting polymer structures has the potential to store sensitive information within a material and achieve hardware-level information security. 

\section*{Acknowledgments}
The authors thank the French National Nanofabrication Network \href{https://www.renatech.org/en/}{RENATECH} for financial support of the IEMN cleanroom. 
We thank also the IEMN cleanroom staff for their advice and support. 
This work is funded by ANR-JCJC "Sensation" project (grant number: \href{https://anr.fr/Projet-ANR-22-CE24-0001}{ANR-22-CE24-0001}), the ERC-CoG "IONOS" project (grant number: \href{https://doi.org/10.3030/773228}{773228}) and the R{\'e}gion Hauts-de-France.

\section*{Competing Interests}
The authors declare no competing interests.

\bibliographystyle{natsty-doilk-on-jour}  
\bibliography{ref}  

\newpage

\section*{Supplementary Materials: On Electropolymerized Fingerprints and their Potential for Identification and Encryption}

\beginsupplement

\begin{figure}[!h]
  \centering
  \includegraphics[width=0.95\columnwidth]{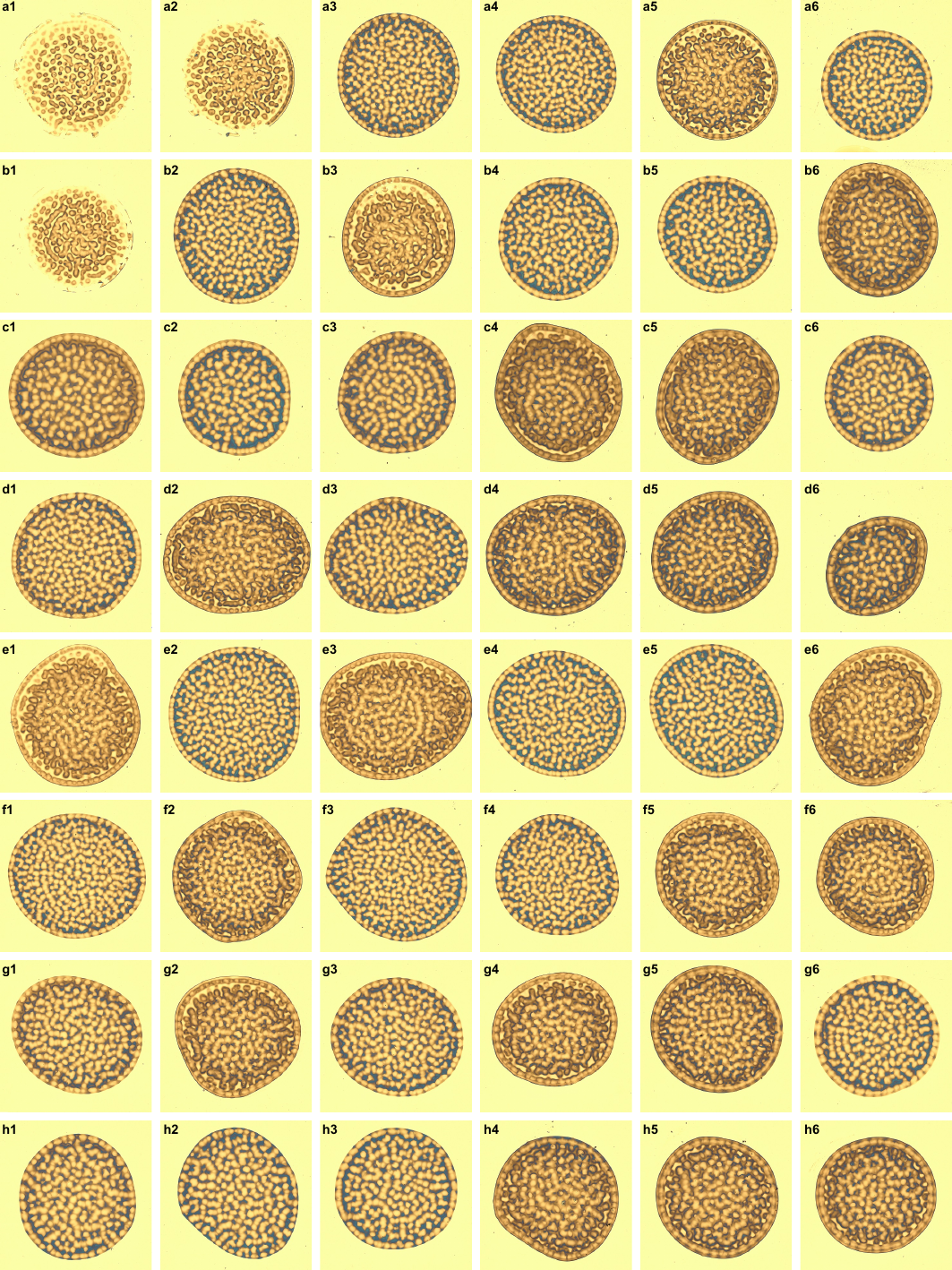}
  \caption{\textbf{Microscope picture for the 2$\times$24 patterns generated from both solutions $\vert$ }
  \textbf{a,} 
  All depositions were realized on a same substrate with six iterations of parallel electropolymerizations from \textbf{a.} to \textbf{f.} in chronological order.
  To prepare one electropolymerization, solution droplets were manually deposited with a pipette from \textbf{.1} to \textbf{.6}, in chronological order.
  } 
  \label{fig:figS1}
\end{figure}

\end{document}